\documentclass[superscriptaddress,showpacs,aps,twocolumn,prb,floatfix]{revtex4}
\usepackage{bm,amsmath,amssymb}
\usepackage{color}

\usepackage[dvips]{graphicx}

\begin{document}

\title{Width of the charge-transfer peak in the SU($N$) impurity Anderson
model and its relevance to non-equilibrium transport}
\author{J.~Fern\'andez}
\affiliation{Centro At\'{o}mico Bariloche and Instituto Balseiro, Comisi\'{o}n Nacional
de Energ\'{\i}a At\'{o}mica, CONICET, 8400 Bariloche, Argentina}
\author{F.~Lisandrini}
\affiliation{Instituto de F\'{\i}sica Rosario. Facultad de Ciencias Exactas Ingenier\'{\i}a 
y Agrimensura, Universidad Nacional de Rosario. Bv. 27 de Febrero 210 bis,
2000 Rosario, Argentina}
\author{P.~Roura-Bas}
\affiliation{Dpto de F\'{\i}sica, Centro At\'{o}mico Constituyentes, Comisi\'{o}n
Nacional de Energ\'{\i}a At\'{o}mica, CONICET, Buenos Aires, Argentina}
\author{C.~Gazza}
\affiliation{Instituto de F\'{\i}sica Rosario. Facultad de Ciencias Exactas Ingenier\'{\i}%
a y Agrimensura, Universidad Nacional de Rosario. Bv. 27 de Febrero 210 bis,
2000 Rosario, Argentina}
\author{A.~A.~Aligia}
\affiliation{Centro At\'{o}mico Bariloche and Instituto Balseiro, Comisi\'{o}n Nacional
de Energ\'{\i}a At\'{o}mica, CONICET, 8400 Bariloche, Argentina}
\date{\today }

\begin{abstract}
We calculate the width $2\Delta_{\text{CT}}$ and intensity of the charge-transfer peak 
(the one lying at the on-site energy $E_d$) in the impurity
spectral density of states as a function of $E_d$ in the SU($N$) impurity
Anderson model (IAM). We use the dynamical density-matrix renormalization group (DDMRG)
and the noncrossing-approximation (NCA) for $N$=4, and a 1/$N$ variational
approximation in the general case. In particular, while for $E_d \gg \Delta$, 
where $\Delta$ is the resonant level half-width, $\Delta_{\text{CT}}=\Delta$ as
expected in the noninteracting case, for $-E_d \gg N \Delta$ one has 
$\Delta_{\text{CT}}=N\Delta$. In the $N$=2 case, some effects of the variation of $%
\Delta_{\text{CT}}$ with $E_d$ were observed in the conductance through a quantum
dot connected asymmetrically to conducting leads at finite bias [J.
K\"onemann \textit{et al.}, Phys. Rev. B \textbf{73}, 033313 (2006)]. More
dramatic effects are expected in similar experiments, that can be carried
out in systems of two quantum dots, carbon nanotubes or other, realizing the
SU(4) IAM.
\end{abstract}

\maketitle

%73.21.La Quantum dots (electron states and collective excitations
%73.63.Kv Quantum dots (transport)
%73.63.-b Electronic transport in nanoscale materials and structures	 
%73.22.-f Electronic structure of nanoscale materials: clusters, nanoparticles, nanotubes, and nanocrystals
%75.20.Hr Local moment in compounds and alloys; Kondo effect, valence	 fluctuations, heavy fermions 
%72.15.Qm Scattering mechanisms and Kondo effect)

\section{Introduction}

\label{intro}

The discovery of the Kondo effect \cite{hewson} in semiconducting quantum
dots (QDs),\cite{gold,cro,gold2,wiel,grobis,kreti,ama,hubel,keller} has
spurred the study of electronic transport through QDs. Later molecular 
QDs,\cite{liang,kuba,yu,leuen,parks,roch,scott,parks2,NatelsonReview,serge,vincent}
and QDs in carbon nanotubes,\cite{jari,maka,anders} were also studied.
Molecular QDs in general have large asymmetric coupling to source and drain
leads. Semiconducting QDs are characterized by the enormous possibilities
for tuning the different parameters. In all these systems, several physical
effects are generically observed when the system is cooled at cryogenic
temperatures due to the large Coulomb repulsion in these nanoscopic QDs,
such as Coulomb blockade and the Kondo effect, which implies a resonance at
the Fermi energy in the spectral density of the dot state, that leads to an
anomalous peak in the differential conductance $G(V)=dI/dV$ at zero bias
voltage $V$, where $I$ is the current through the QD. These physical effects are
usually well described by an impurity Anderson model (IAM).

While intense research has been devoted to the Kondo effect, also important
effects of the interactions take place at finite $V$ that are unexpected for
independent electrons. In particular, Coulomb blockade peaks were shown to
present a strong width renormalization in the situation of large tunnel
asymmetries between source and drain electrodes.\cite{haug} Specifically, K%
\"{o}nemann \textit{et al.} started from an equilibrium situation in which
the dot level $E_{d}>0$. Then they applied a bias voltage, which shifts the
chemical potential $\mu _{L}$ of one of the leads (we call it the left lead)
in $eV>0$ and also $E_{d}$ by about $eV/2$ due to capacitance effects. A
peak in $G(V)$ is observed when $\mu _{L}\sim E_{d}$. When instead a bias
voltage such that $eV<0$ is applied, another peak in $G(V)$ was observed
(when $E_{d}\sim \mu _{R}$, the chemical potential of the right lead). In a
noninteracting picture, one would expect that both peaks have nearly the
same width and the same height.\cite{park,capac} However, the latter turns
out to be about two times wider and with a maximum nearly five times smaller.\cite{haug} 
The reason is that the QD has a much stronger coupling to the
right lead, and then for $eV>0$ the QD is empty, while for $eV<0$ it is in
an intermediate valence situation, and one knows that the half width at half
maximum $\Delta _{\text{CT}}$ of the spectral density of the charge-transfer (CT) peak
in the IAM increases with the occupancy.\cite{capac,pru,logan} An
explanation of the experiment in the framework of the SU(2) IAM, including
the effects of capacitance and tunneling asymmetries and the Kondo
effect is provided in Ref. \onlinecite{capac}.

Several tunable systems of the kind discussed above in which there is an
orbital, dot or valley degeneracy, in addition to the spin one, are
described by the SU(4) IAM, like those of carbon nanotubes,\cite{jari,maka,anders} 
a nanoscale Si transistor,\cite{tetta} an As dopant in a
Si nanostructure,\cite{lans} and systems of two QDs where the occupation of
one or the other QD plays the role of the orbital degeneracy.\cite{hubel,keller} 
In the latter case, the tunneling coupling of both dots are
in general different, reducing the symmetry to SU(2), but tuning other
parameters the SU(4) symmetry can be recovered as an emergent one at low
temperatures.\cite{restor,nishi}. Crossovers 
\cite{tetta,buss,desint,buss2,thermo,nishi2,oks,lopes} and abrupt transitions \cite{klee} between
SU(2) and SU(4) symmetry were also studied. 

In the noninteracting case, for a flat very wide conduction band as we assume,
the spectral density of the localized level for each of the N components 
in the SU($N$) impurity Anderson model is a Lorentzian with half width at half maximum 
$\Delta$ [defined by Eq. \ref{delta} in terms of the microscopic parameters of the model]. 
Thus in this case, $\Delta _{\text{CT}}=\Delta$ independently of on site energy $E_d$.
As we show in this work, the
effect of correlations on $\Delta _{\text{CT}}$ increases with N and one expects a
width $\Delta _{\text{CT}}=N\Delta $ for the SU($N$) model in the Kondo regime $-E_d, E_d+U \gg \Delta$,
where $U$ is the Coulomb repulsion.
Therefore if the
experiments like those of K\"{o}nemann \textit{et al.} are performed for any
of the above realizations of the SU(4) IAM, this effect would be more
evident. Furthermore for the experiments with the setup of two QDs in which
each QD is connected to its own pair of conducting leads and the tunneling
matrix elements and the voltages at the four leads can be controlled
independently with high precision,\cite{ama,keller} the spectral density can
be read out directly form the differential conductance through one of the QD
under appropriate conditions.\cite{oks}

In this work we use a 1/N expansion based on variational wave functions,\cite{varma,gunn} 
to show that one expects in general that $\Delta_{\text{CT}} \sim N
\Delta$ for $-E_d \gg N \Delta$ in the SU($N$) IAM. We also calculate $\Delta_{\text{CT}}$ 
as a function of $E_d$ for N=4 using dynamical density-matrix
renormalization group (DDMRG) and the noncrossing-approximation (NCA). In
this case $\Delta_{\text{CT}}$ changes rather abruptly from $4 \Delta$ to $\Delta$
as the effective $E_d$ (including renormalization due to effects of the
hybridization \cite{hald}) changes sign from negative to positive. We also
discuss the possible experimental relevance of these results.

The paper is organized as follows. In Sec. \ref{model}, we describe the  SU($N$) 
IAM and the methods used. In Sec. \ref{vari} we discuss the CT peak 
for general $N$ at zero temperature in the Kondo regime $-E_{d}\gg N\Delta $ 
using a variational approximation. 
Sec. \ref{su4} contains NCA and DDMRG results for $N$=4.
In Sec. \ref{occ} we show the results for the occupancy as a function of on-site energy level 
and compare them with the noninteracting case and with exact Bethe ansatz results.
In Sec. \ref{exp} we 
discuss the relevance of our results to possible experimental realizations. 
Sec. \ref{sum} contains a summary and a discussion.

\section{Model and methods}

\label{model}

We consider the one-level SU($N$) IAM with infinite on-site repulsion $U$. The
impurity states involve a singlet configuration $|s\rangle $ together
with a degenerate configuration $|m\rangle $, $m=1$ to N, corresponding to
one additional electron (or hole) in the \textquotedblleft
impurity\textquotedblright , which can be a QD, a system of two QDs, a part
of a carbon nanotube, or an atom with degenerate levels as discussed in the
introduction. When discussing transport experiments, for simplicity we
assume that the \textquotedblleft impurity\textquotedblright\ is connected
to a left ($L$) and a right ($R$) conducting lead. An extension to the case
of the system of two QDs in which both are connected to a pair of
independent leads \cite{ama,hubel,keller} is straightforward.\cite{oks} The
Hamiltonian reads: 
\begin{eqnarray}
H &=&\sum_{m}E_{d}|m\rangle \langle m|+\sum_{\nu km}\epsilon _{\nu k}c_{\nu
km}^{\dagger }c_{\nu km}  \notag \\
&&+\sum_{\nu km}(V_{k}^{\nu }|m\rangle \langle s|c_{\nu km}+\mathrm{H.c}.),
\label{ham}
\end{eqnarray}
where the constraint $|s\rangle \langle s|+\sum_{m}|m\rangle \langle m|=1$
is imposed (this is equivalent to the assumption of infinite Coulomb
repulsion $U$). Here $c_{\nu km}^{\dagger }$ create conduction states in the
lead $\nu $ with projection $m$ and wave vector $k$. The tunnel couplings of
the quantum dot to the leads and the total resonant level widths are

\begin{eqnarray}
\Delta _{\nu }&=&\pi \sum_{k}|V_{k}^{\nu}|^{2}\delta (\omega -\epsilon _{\nu k}), \notag \\
\Delta &=&\Delta _{L}+\Delta _{R}
\label{delta}
\end{eqnarray}
taken in general independent of energy $\omega $.\cite{note} For the discussion of the impurity spectral density $\rho
_{m}(\omega )$, only $\Delta $ is relevant

To obtain $\rho _{m}(\omega )$ we use three different methods. The
variational one, based on Refs. \onlinecite{varma,gunn} is the simplest one
and is described in detail in the next Section. It is limited to the Kondo
regime $-E_{d}\gg N\Delta $ and zero temperature. We also use NCA and DDMRG. 

The NCA is equivalent to a sum of an infinite series of diagrams in
perturbations in the hybridization.\cite{hewson,bickers,kroha} In the Kondo
regime, it is known to reproduce correctly the relevant energy scale $T_{K}$
and its dependence of parameters. An advantage for our purpose over the
numerical-renormalization group in which finite-energy features are
artificially broadened due to the logarithmic discretization of the
conducting band,\cite{zitko,loig} NCA correctly describes these features.
For instance, the NCA works satisfactorily in cases in which the conduction
density of states is not smooth,\cite{kroha} including in particular a step
in the conduction band.\cite{ds} Furthermore, it has a natural extension to
non-equilibrium conditions,\cite{wingreen} and it is specially suitable for
describing satellite peaks of the Kondo resonance, as those observed in Ce
systems,\cite{reinert,ehm} or away from zero bias voltage in non-equilibrium
transport due to phonons \cite{fon,sate} or magnetic and orbital excitations.\cite{restor,NFL,st} 
Alternatives to NCA for non-equilibrium
problems are renormalized perturbation theory (but limited to small $\omega $, 
$V$ and temperature $T$)\cite{hbo,ng,com3} or the equation of motion
method,\cite{rome1,rapha} although it does nor reproduce correctly the
functional dependence of $T_{K}$ on $E_{d}$.\cite{rapha,rome2}

However, the NCA has important limitations out of the Kondo regime. In
particular for moderate positive $E_{d}$, the impurity self energy has an
unphysical positive imaginary part (this means that the Green function
violates causality) and the impurity spectral density presents a spurious
peak at the Fermi energy. As a consequence, the NCA results in this region of $E_{d}$ are unreliable. 
For this reason, we also use the DDMRG in its correction-vector-method approach. Since it was 
introduced by Kuehner and White,\cite{kuhne99a} 
the correction vector has shown to be a reliable way to do dynamical calculations with DMRG in 
different low-dimensional strongly correlated models. Different strategies were done to 
include the correction vector in the target states of a standard DDMRG.\cite{hovel00,freun93a,jecke02} 
We choose a recently presented version introduced by Nocera and 
Alvarez,\cite{alva} which based on a Krylov-space approach, has been shown to 
be more accurate and efficient than conjugate gradient.\cite{jecke02}

However, the two major difficulties of the correction-vector method persist. First, the need to be 
computed in small frequency intervals which is unavoidable but with parallelization strategies becomes affordable.
Second the artificial broadening $\eta$ that is necessarily introduced in the calculation 
(calculations cannot be done at $\eta=0$). Therefore one is  computing $G(\omega+i\eta)$, and the
resulting
impurity density spectral function $\rho^{(\eta)}(\omega) = - \frac{1}{\pi} {\text{Im}} G(\omega+i\eta)$ 
can be visualized as the real spectral density $\rho(\omega)$ convoluted by the Lorentzian 
$\rho_d(\omega) =(\eta/\pi)/((\omega-\omega_0)^2+\eta^2)$ of width $\eta$. 
Since we are interested in the real spectral density for $\eta \rightarrow 0^+$, 
this compels to a deconvolution of the spectrum, and 
many proposals have been presented to do this difficult task 
successfully,\cite{Gebhard2003,Ulbricht2010,Weichselbaum2009,Nishimoto2004,Raas2004,Raas2005,Paech1014}
and resolve structures with small width in the real spectrum.
Fortunately, as it was mentioned above, we are interested in the line shapes of the CT peak for the IAM
which as it has been very well described in Ref. \onlinecite{Raas2004}, the deconvolution becomes easiest,  
particularly for large $U$. The authors have also shown that the noninteracting case $U=0$ is well described by the 
DDMRG.\cite{Raas2004}

\begin{figure}[h]
\includegraphics[width=8.cm]{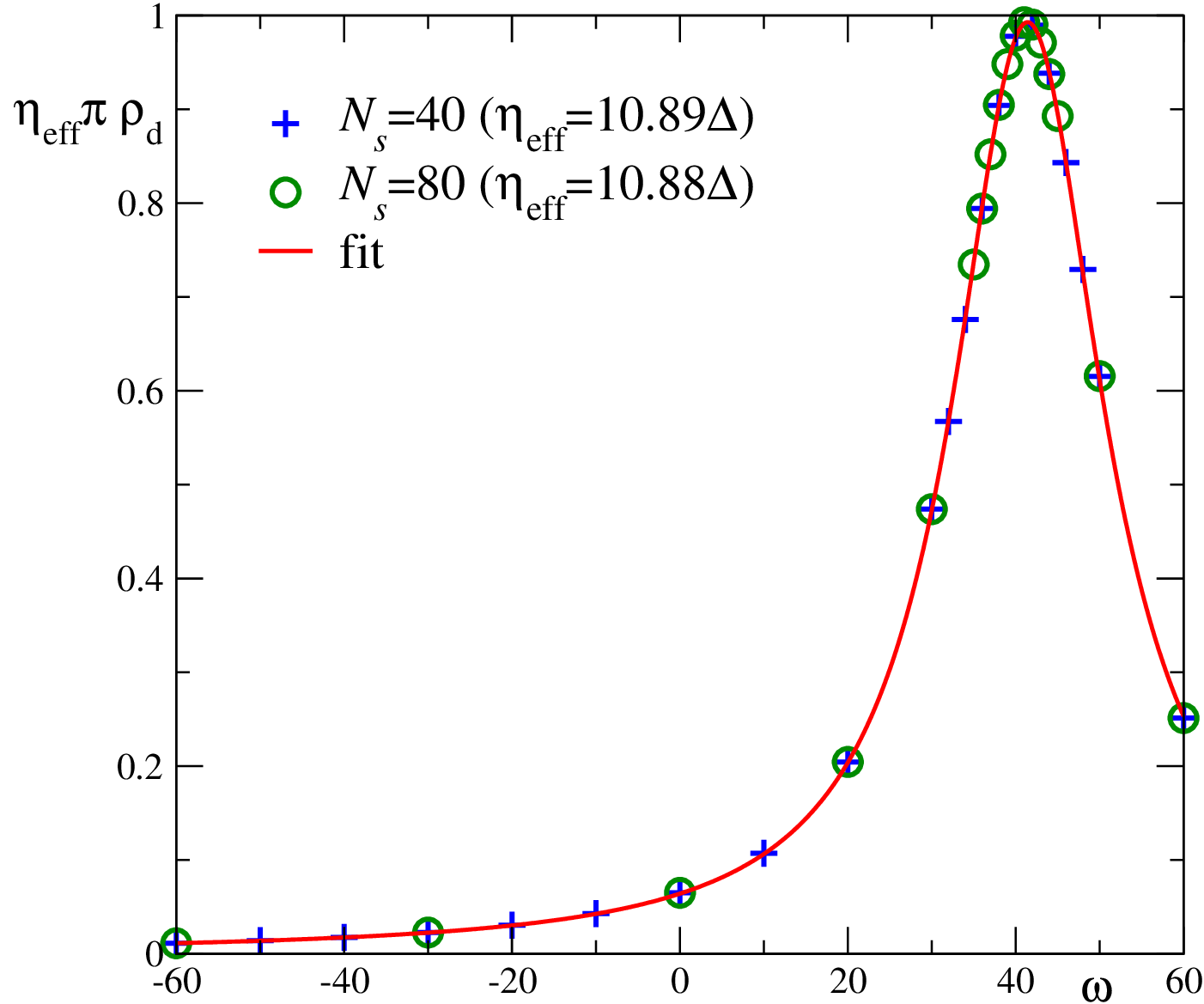}
\caption{(Color online) DDMRG correction vector data of the CT at $\eta=10\Delta$. 
Open circles (Crosses) for $N_s=40$ ($N_s=80$), for $E_d=40\Delta$ (see Fig.~\ref{anchos}). 
In legends we indicate the $\eta_{\text{eff}}$ obtained from a fit with a Lorentzian   
$\propto K/((\omega-\omega_{up})^2+\eta_{\text{eff}}^2)$ (thin line) for each size.}
\label{eta}
\end{figure}

Assuming that the original CT peak is a Lorentzian, then the data obtained with DDMRG will be
the convolution of two Lorentzians of half width at half maximum $\Delta_{\text{CT}}$ and $\eta$ respectively. 
This convolution will
result in a new Lorentzian whose half width is the sum of the two original ones, so that the half width 
at half maximum
of the peak obtained by fitting the DDMRG calculations will be $\eta_{\text{eff}}=\Delta_{\text{CT}}+\eta$.
The strategy works fine while the CT peak does not merge with the Kondo peak, obtaining reliable 
results in the range $E_d \ge 10\Delta$. When both peaks overlap, we still can distinguish the left part of the CT 
(at lower frequencies) and we fit just the left half part of CT peak using the DDMRG data.
Calculations were done with up to 1000 states keeping the truncation error less than $10^{-8}$, 
assuring a numerical error in DDMRG data much smaller than the size of the symbols in Fig. \ref{eta}. 

\vspace{0.5cm}

In order to visualize the procedure, we show in Fig.~\ref{eta} the DDMRG correction vector results, 
for $E_d=40\Delta$ with $\eta=10\Delta$, and for two sizes ($N_s=40$, 80). As it is apparent in the figure, 
there are no significant finite-size effects. 
Then we fit the peaks with Lorentzians to obtain the half width $\eta_{\text{eff}}$ and the position 
$E_{d}^{\text{eff}}$ for each lattice size, allowing us to take the estimation $\eta_{\text{eff}}=(10.9\pm0.1)\Delta$. 
Finally we can calculate the values presented in Fig.~\ref{anchos} 
doing the subtraction of the width obtained with the $\eta=10\Delta$ used in the correction vector. 
We derive in our example $\Delta_{\text{CT}}=(0.9\pm0.1)\Delta$.

\section{Width and weight of the charge-transfer peak for large negative on-site energy}

\label{vari}

In this Section we calculate $\Delta _{\text{CT}}$ well inside the Kondo regime $%
-E_{d}\gg N\Delta $, using a variational procedure.\cite{varma,gunn}

The ground-state wave function of the Hamiltonian Eq. (\ref{ham}) is
approximated as

\begin{equation}
|g\rangle =A|F\rangle +\sum_{qm}B_{q}d_{m}^{\dagger }c_{qm}|F\rangle ,
\label{gvar}
\end{equation}%
where $|F\rangle =\Pi _{q}^{\prime}c_{qm}^{\dagger }|0\rangle \bigotimes |s\rangle$ 
is the state with the filled
Fermi see in the band and the impurity in the singlet state, the subscript 
$q $ refers to both lead and wave vector index [$\nu $ and $k$ in Eq. (\ref%
{ham})] , the prime over the product symbol means that only $q$ for which $%
\epsilon _{q}<\epsilon _{F}$ are included, where $\epsilon _{F}$ is the
Fermi energy. We denote $d_{m}^{\dagger }=|m\rangle \langle s|$. The
variational parameters $A$ and $B_{q}$ are determined minimizing the energy.
The $B_{q}$ are independent of $m$ as a consequence of SU($N$)\ symmetry.
Normalization of $|g\rangle $ implies

\begin{equation}
|A|^{2}+N\sum_{q}|B_{q}|^{2}=1.  \label{norm}
\end{equation}%
In the following we take the origin of energies at the Fermi energy $%
\epsilon _{F}=0$. We define the energy gain due to hybridization as 

\begin{equation}
T_{K}=E-E_{0},
\label{tkvar1}
\end{equation}

where $E$ is the ground-state energy and $E_{0}=E_{d}+\langle F|H|F\rangle $, 
the ground-state energy for $V_{q}=0$.
Minimizing the energy one obtains

\begin{equation}
B_{q}=\frac{-\bar{V}_{q}A}{T_{K}-\epsilon _{q}},  \label{bq}
\end{equation}%
and the equation for the ground-state energy, which can be written in the
form

\begin{equation}
T_{K}-E_{d}-N\sum_{q}\frac{|V_{q}|^{2}}{T_{K}-\epsilon _{q}}=0.  \label{evar}
\end{equation}%
The sum can be evaluated assuming $\Delta =\Delta _{L}+\Delta _{R}=\pi
\sum_{q}|V_{q}|^{2}\delta (\omega -\epsilon _{q})$ independent of $\omega $
in the range $-D<\omega <D$ and gives $(\Delta /\pi )\ln [(D+T_{K})/T_{K}]$.
In the Kondo limit $-E_{d}\gg N\Delta $, one can neglect $T_{K}$ in
comparison with $D$ and $|E_{d}|$ obtaining

\begin{equation}
T_{K}=D\exp \left( \frac{\pi E_{d}}{N\Delta }\right) ,  \label{tkvar}
\end{equation}%
which has the correct exponential dependence on $E_{d}$ (although the
correct prefactor is smaller.\cite{hewson,fili})

At zero temperature, the impurity spectral density is

\begin{eqnarray}
\rho _{m}(\omega ) &=&\rho _{m}^{d}(\omega )+\rho _{m}^{c}(\omega ),  \notag
\\
\rho _{m}^{d}(\omega ) &=&\sum_{e}|\langle e|d_{m}|g\rangle |^{2}\delta
(\omega +E_{e}-E),  \notag \\
\rho _{m}^{c}(\omega ) &=&\sum_{e}|\langle e|d_{m}^{\dagger}|g\rangle
|^{2}\delta (\omega -E_{e}+E),  \label{rhov}
\end{eqnarray}%
where $E_{e}$ is the energy of the excited state $|e\rangle $. Since we are
interested in $\omega $ near $E_{d}$ and in this Section $E_{d}$ is well
below the Fermi energy, we can neglect the creation part $\rho
_{m}^{c}(\omega )$ of the spectral density. Using Eq. (\ref{gvar}) we can
write

\begin{eqnarray}
\rho _{m}^{d}(\omega ) &=&-\frac{1}{\pi }\sum_{q}|B_{q}|^{2}\text{Im}%
G_{qm,qm}(E-\omega ),  \notag \\
G_{qm,qm}(z) &=&\langle qm|\hat{G}|qm\rangle ,\text{ }|qm\rangle
=c_{qm}|F\rangle ,  \notag \\
\hat{G}(z) &=&\frac{1}{z+i\eta -H},  \label{gop}
\end{eqnarray}%
where $\eta \rightarrow 0^{+}$ is a positive infinitesimal.

Defining the operators $\hat{G}^{0}(z)=z+i\eta -H_{0}$, where $H_{0}=H-\hat{V%
}$, and $\hat{V}=\sum_{qm}(V_{q}d_{m}^{\dagger}c_{qm}+\mathrm{H.c}.)$ is
the hybridization part of $H$, one has the Dyson equation

\begin{equation}
\hat{G}=\hat{G}^{0}+\hat{G}^{0}\hat{V}\hat{G}.  \label{dyson}
\end{equation}%
Taking matrix elements of both members Eq. (\ref{dyson}) between states $%
|qm\rangle $ and/or $|q^{\prime}m^{\prime}qm\rangle =d_{m^{\prime}}^{\dagger
}c_{q^{\prime}m^{\prime}}|qm\rangle $, we obtain after some algebra

\begin{eqnarray}
&&G_{qm,qm}(z)= \notag \\
&&=\frac{G_{qm,qm}^{0}(z)}{1-G_{qm,qm}^{0}(z)\sum_{q^{\prime
}m^{\prime }}|V_{q^{\prime }}|^{2}G_{q^{\prime }m^{\prime }qm,q^{\prime
}m^{\prime }qm}^{0}(z)},  \notag \\
&&G_{qm,qm}^{0}(z)=\frac{1}{z+i\eta +\epsilon _{q} -E_{0}+E_{d}},  \notag \\
&&G_{q^{\prime }m^{\prime }qm,q^{\prime }m^{\prime }qm}^{0}(z)=\frac{1}{%
z+i\eta +\epsilon _{q}+\epsilon _{q^{\prime }}-E_{0}}.  \label{gqm}
\end{eqnarray}%
The sum over $m^{\prime }$ above is just a factor $N$ due to SU($N$) symmetry.
As we shall see, this factor enters into the width of the charge-transfer (CT)
peak $\Delta _{\text{CT}}$ The sum over $q^{\prime }$ can be evaluated using the
density of conduction states [as we have done to obtain Eq. (\ref{tkvar})],
leading to

\begin{eqnarray}
\text{Im}G_{qm,qm}(E-\omega ) &=&\frac{-N\Delta }{\left( \omega
+T_{K}-\epsilon _{q}-E_{d}-\Lambda \right) ^{2}+\left( N\Delta \right) ^{2}},
\notag \\
\Lambda &=&\frac{N\Delta }{\pi }\ln \left\vert \frac{D+\omega
+T_{K}-\epsilon _{q}}{\omega +T_{K}-\epsilon _{q}}\right\vert ,
\label{imgqm}
\end{eqnarray}

Inserting this in the first Eq. (\ref{gop}) and using Eqs. (\ref{norm}) and 
(\ref{bq}) one obtains the desired spectral density.

In the limit $-E_{d}\gg N\Delta $, $T_{K}$ becomes exponentially small [see
Eq. (\ref{tkvar})]. Furthermore, from Eq. (\ref{bq}), one realizes that the
values of $\epsilon _{q}$ which lead to the larger values of $B_{q}\sim
A/T_{K}$ are of the order of $T_{K}$. From Eqs. (\ref{norm}) and (\ref{bq})
one sees that for $T_{K}\rightarrow 0$, also $A\rightarrow 0$ and $%
\sum_{q}|B_{q}|^{2}\rightarrow 1/N$. Thus in this limit $%
T_{K}\rightarrow 0$, for any function $F(\epsilon _{q})$, we can use

\begin{equation}
\sum_{q}|B_{q}|^{2}F(\epsilon _{q})\rightarrow \frac{1}{N}F(0)  \label{lim}
\end{equation}%
Furthermore since $\Lambda $ has a week logarithmic dependence and we are
interested in $\omega \sim E_{d}$, we can evaluate it at $E_{d}$. Then Eqs. (%
\ref{imgqm}), the first (\ref{gop}) and (\ref{lim}) lead to

\begin{eqnarray}
\rho _{m}(\omega ) &\simeq &\frac{\Delta /\pi }{\left( \omega -E_{d}^{\text{%
eff}}\right) ^{2}+\left( N\Delta \right) ^{2}},  \label{rovar} \\
E_{d}^{\text{eff}} &=&E_{d}+\frac{N\Delta }{\pi }\ln \left\vert \frac{D+E_{d}%
}{E_{d}}\right\vert ,  \label{edeff}
\end{eqnarray}%
well inside the Kondo limit $-E_{d}\gg N\Delta .$

This simple result reflects three important physical effects of the
correlations on the CT peak: i) The position is shifted upwards. The origin of this shift
was explained by Haldane for the SU(2) case,\cite{hald} and extensions to
the SU(4) case and the SU(2) $\leftrightarrow $SU(4) crossover were
discussed.\cite{fili,restor} Depending on details different but similar
estimations for the shift were given. This point is discussed in more detail in Section 
\ref{t0} (see Fig. \ref{shift}). ii) The half width at half maximum of
the peak is $\Delta _{\text{CT}}=N\Delta $. This is a factor $N$ with respect to
the noninteracting case. iii) the total weight of the peak is $1/N$. The
maximum intensity is thus reduced in a factor $1/N^{2}$ compared to the
noninteracting case.

\section{Width of the charge-transfer peak in the SU(4) IAM}

\label{su4}

In this section we show our results for $\Delta_{\text{CT}}$ as a function of $E_d$
in the SU(4) case using NCA and DDMRG. For the NCA we used a constant density of 
unperturbed conduction states extending from $-D$ to $D$ with $D=100 \Delta$.
The assumptions made in DDMRG imply a semielliptical density of conduction 
states.\cite{note,proe}

\subsection{Temperature $T \rightarrow 0$}
\label{t0}

\begin{figure}[h]
\includegraphics[width=8.cm]{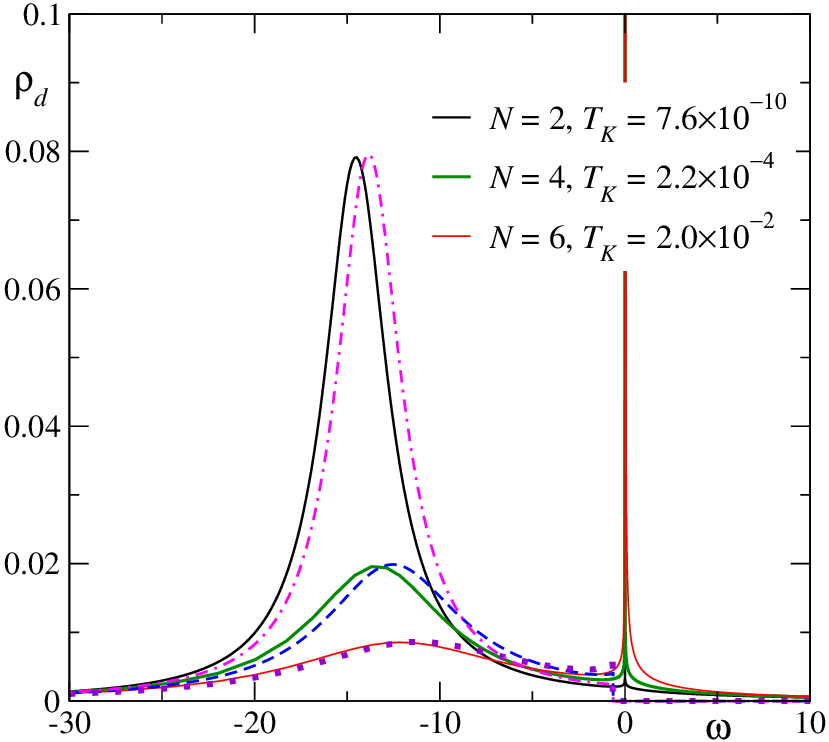}
\caption{(Color online) 
Impurity spectral density calculated with
the NCA (full lines) and variational wave function (dashed lines) for $E_d= - 15 \Delta$. }
\label{nca}
\end{figure}

We discuss first the results at low temperature ($T = T_{K}/10$). Here $T_{K}$ 
was chosen such that $G(T_{K})=G_{0}/2$, where $G_{0}$ is the conductance of the system at $T=0$ 
in the extreme Kondo limit in which the total occupancy is 1 (the maximum possible conductance of the system).
Using Friedel sum rule \cite{lang,yoshi} one has 

\begin{equation}
G_0 = N \frac{e^2}{h} \sin ^{2}\left( \frac{\pi}{N} \right).
\label{g0}
\end{equation}

In Fig.~\ref{nca} we show the spectral density $\rho_m(\omega)$ obtained with the NCA 
in the Kondo regime ($-E_{d}\gg N\Delta $). 
%and the empty orbital one ($E_{d}\gg \Delta $). In the former case, $\rho_d(\omega)$ 
It shows two peaks. The broad CT peak
(which is the focus of this work) at $E_{d}^{\text{eff}}$ (slightly larger than $E_{d}$)
and the very narrow Kondo peak at the Fermi energy. Instead, for positive $E_{d}$ (not shown)
only the CT peak is present. We also show in the figure the result for 
$\rho _{m}^{d}(\omega )$ obtained using Eq. (\ref{gop}), (\ref{norm}) and (\ref{bq}).
One can see that there is a very good agreement between the NCA and variational 
calculations for the CT peak. 
We have verified that the CT peak in the empty orbital regime (not shown) looks identical to the 
noninteracting spectral density, as expected from the discussion of the previous section. 
Instead, in the Kondo regime it is nearly four times broader and its maximum is sixteen times smaller.

We have repeated these calculations for several values of $E_d$. In the empty
orbital regime, the peak has been fit with a Lorentzian with three parameters:
the width identified with $\Delta _{\text{CT}}$, the position $E_{d}^{\text{eff}}$ 
and its weight $w$, which is the integral in energy of the Lorentzian.
In the Kondo regime we used a similar fit with two Lorentzians, one for the CT 
peak and one for the Kondo peak. In the intermediate valence (IV) regime 
$|E_{d}^{\text{eff}}| \sim \Delta$, this procedure failed because in addition to the fact 
that the CT peak merges with a Kondo peak that becomes increasingly wider,
the NCA fails and gives a self energy with a positive imaginary part.
Thus, in this region, the DDMRG results are particularly useful. 
%A similar criterion of fitting with two Lorenztians in the Kondo regime has been used for the DMRG results.

\begin{figure}[h]
\vspace{1.cm}
\includegraphics[width=8.cm]{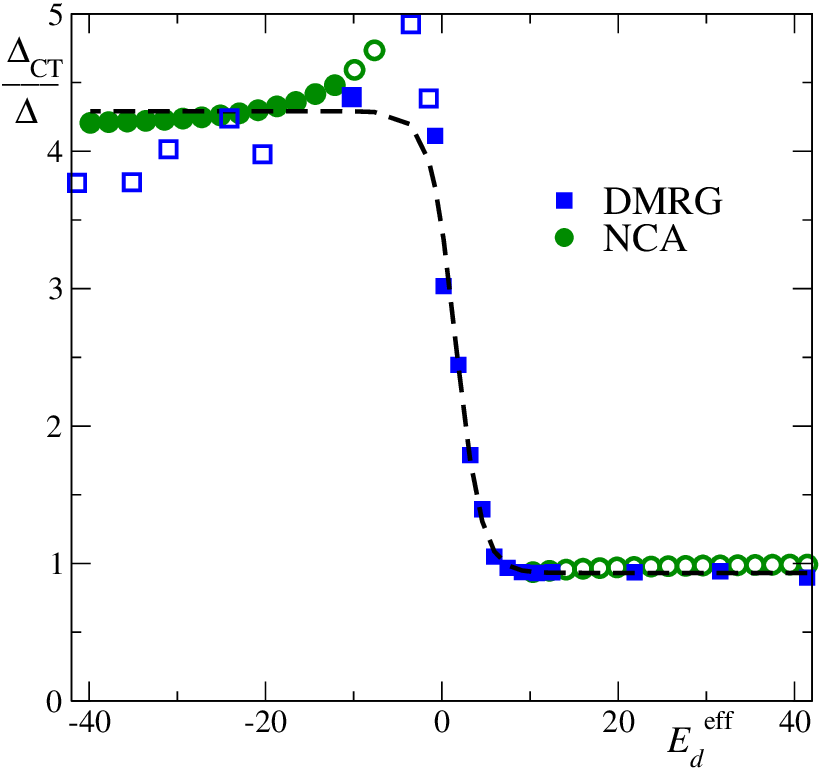}
\caption{(Color online) Half width at half maximum of the charge transfer peak as a function of
the effective on site energy using NCA (green circles) and DDMRG (blue squares). 
Dashed lines is a fit of the filled symbols using Eq. (\ref{dfit}).}
\label{anchos}
\end{figure}

In Fig.~\ref{anchos} we show the resulting $\Delta _{\text{CT}}$ as a function of $E_{d}^{\text{eff}}$ 
using both techniques.
The filled symbols denote the most reliable result between both.
Due to the fact that the band is flat in the NCA and semielliptical in DDMRG (with a total width 
$2D=200 \Delta$), the resonant level width
$2\Delta_{\text{DDMRG}}(\omega)$ is energy dependent in DDMRG and smaller than
the corresponding width $2\Delta$ within NCA, except for $\omega=0$ where both widths are equal.\cite{note}
As a consequence for large $|E_{d}|$, a smaller $\Delta _{\text{CT}}$ is expected in the 
DDMRG calculation, as can be seen in the figure. 
In spite of the lack of NCA results near $E_{d}^{\text{eff}}=0$, the results of both techniques 
indicate a rather sudden crossover from $\Delta _{\text{CT}} \sim 4 \Delta$ to
$\Delta _{\text{CT}} = \Delta$ at a slightly positive $E_{d}^{\text{eff}}$.
In addition, there is an unexpected increase in $\Delta _{\text{CT}}$ for small
negative $E_{d}^{\text{eff}}$. This might be due to the fact that the CT peak is 
deformed as it merges with the Kondo peak in this region.  

For later use (Section \ref{exp}), we also show in the figure a fit with the following function,
which is an extension of that used in the SU(2) case
\begin{equation}
\frac{\Delta _{\text{CT}}}{\Delta }=a-b\tanh \left( \frac{E_{d}^{\text{eff}%
}-E_{\text{IV}}}{c\Delta }\right) .  \label{dfit}
\end{equation}%
For general $N$ one expects $a+b\simeq N$, $a-b\simeq 1$, so that this
expression interpolates between the Kondo regime ($\Delta _{\text{CT}}\simeq
N\Delta $ for $-E_{d}\gg N\Delta $) and the empty orbital one 
($\Delta _{\text{CT}}=\Delta $ for $E_{d}\gg \Delta $) passing through the intermediate
valence (IV) regime.  From the fit we obtain $a=2.61$, $b=1.68$, $c=2.90$,
and $E_{\text{IV}}=1.58\Delta $ for the effective level in the intermediate
valence regime.

\begin{figure}[t]
\includegraphics[width=8.cm]{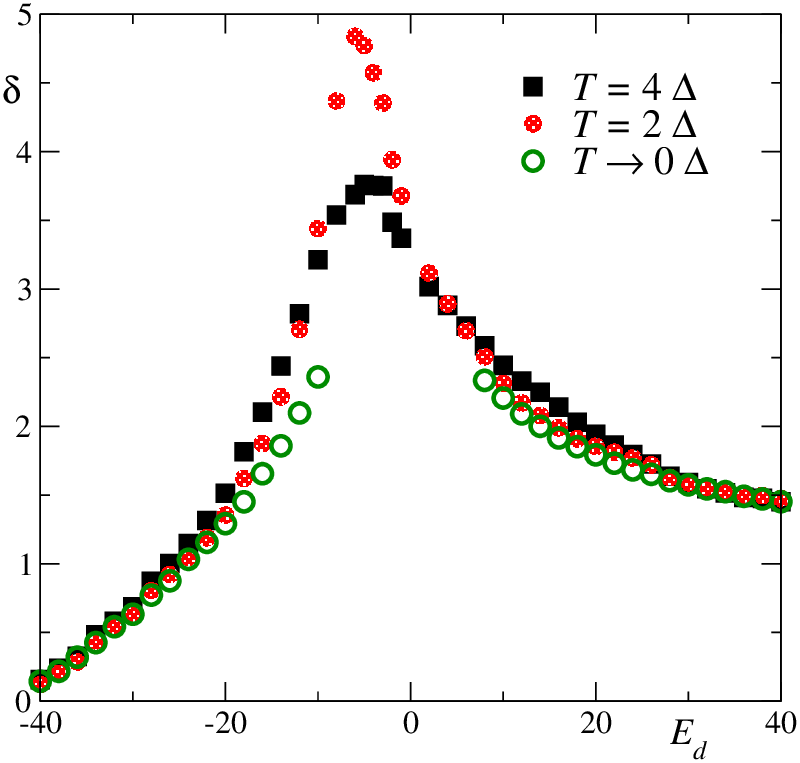}
\caption{(Color online) Shift of the position of the CT peak 
as a function of on-site energy calculated with NCA
for different temperatures $T$.}
\label{shift}
\end{figure}

In Fig.~\ref{shift} we show the shift $\delta = E_{d}^{\text{eff}}- E_d$ as a function of
$E_d$. Approximate expressions of this shift were given based on functional renormalization 
group.\cite{hald,fili,restor} Generalizing Haldane's treatment to the SU($N$) case \cite{restor} one obtains
\begin{equation}
\delta=\frac{N-1}{\pi}\Delta\ln \left(\frac{D}{C}\right),
\label{shifte}
\end{equation}
where $C$ is a low-energy cutoff. Haldane in his study of the SU(2) model used $C=\Delta$, while Filipone {\it et al.} 
have taken $C=|E_d|/\alpha$ with $\alpha$ of the order of 1 in the SU(4) case.\cite{fili}
The NCA results in Fig.~\ref{shift} seem to agree with a cutoff $C$  of the order of the maximum between 
$\Delta$ and $|E_d|$. In the present paper the shift $\delta$ is larger than in similar calculations of 
Ref. \onlinecite{restor}
because of the larger value of $D$ chosen ($D=100$ here and $D=10$ in Ref. \onlinecite{restor}).

\subsection{Finite temperatures}
\label{fint}

\begin{figure}[t]
\includegraphics[width=8.cm]{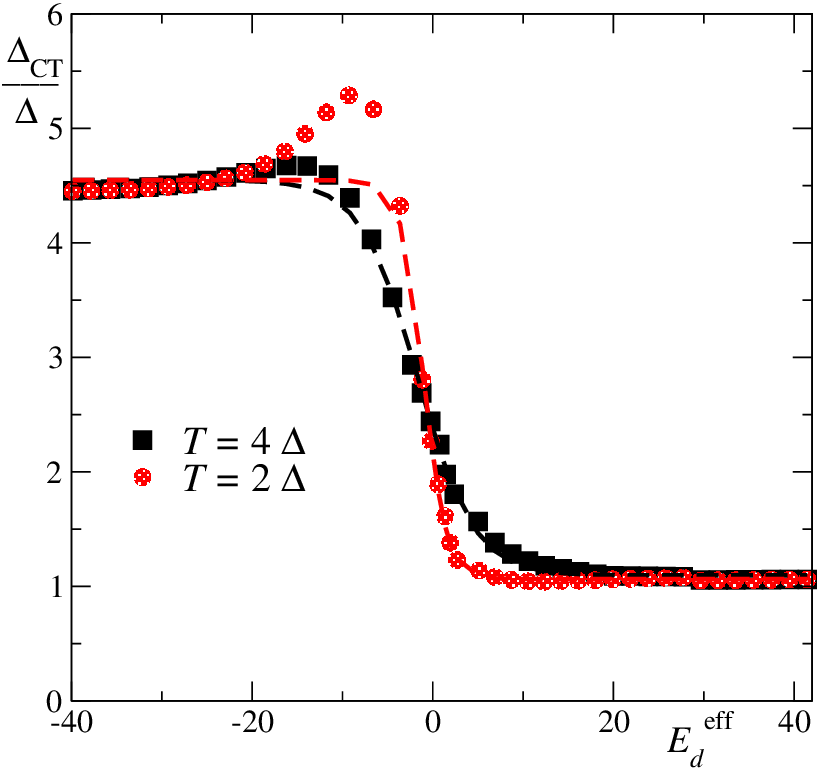}
\caption{(Color online) Half width at half maximum of the charge transfer peak as a function of
the effective on site energy using NCA at two different temperatures $T$. 
Dashed lines are fits using Eq. (\ref{dfit}).}
\label{anchost}
\end{figure}

As the temperature increases, the above mentioned shortcomings of the NCA tend to disappear. 
For $T= 2 \Delta$ the Kondo peak has disappeared (although some weak structure persists near 
$\omega=0$ for small negative $E_{d}^{\text{eff}}$). Thus we have fit the CT peak
using one Lorentzian. The results for $\Delta _{\text{CT}}$ as a function of
$E_{d}^{\text{eff}}$ are plotted in Fig.~\ref{anchost}. For $T= 2 \Delta$ 
still an upturn for slight negative $E_{d}^{\text{eff}}$ is present, as for $T$ near to $0$ (see Fig. \ref{anchos}).
For $T= 2 \Delta$ this structure is greatly reduced and the fit using Eq. (\ref{dfit}) improves.
The parameters of the fit are
for $T= 2 \Delta$: $a=2.81$, $b=1.74$, $c=2.51$, and $E_{\text{IV}}=-0.99\Delta $.  
For $T= 4 \Delta$: $a=2.82$, $b=1.73$, $c=6.26$, and $E_{\text{IV}}=-1.64\Delta $.
Note that between $T= 2 \Delta$ and $T= 4 \Delta$, a broadening of 
crossover region between $\Delta _{\text{CT}} \sim 4 \Delta$ and 
$\Delta _{\text{CT}} \sim  \Delta$ by a factor larger than two takes place.
In addition, between very low $T$ and $T= 2 \Delta$, this region moves from slightly positive 
to slightly negative $E_{d}^{\text{eff}}$.

\section{Occupancy for SU(2) and SU(4)}

\label{occ}

\begin{figure}[t]
\includegraphics[width=8.cm]{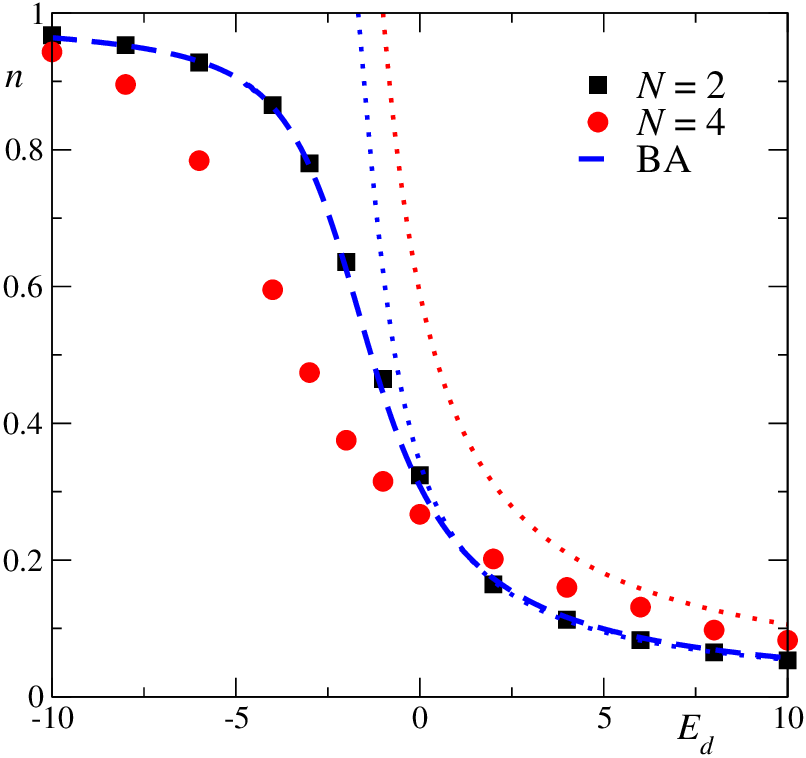}
\caption{(Color online) Total occupancy as a function of on-site energy level calculated with
NCA for the SU($N$) IAM with $N$=2 and 4. Dashed blue line corresponds to the Bethe ansatz (BA) result for 
$N$=2 (see text). Dotted lines are the noninteracting results.}
\label{ocup}
\end{figure}

To illustrate some of the effects of correlations, we show in Fig. \ref{ocup} the total occupancy
$n= \sum_m \langle n_m \rangle$ as a function of the energy level calculated from 
the integral of the pesudofermion density of states in the NCA \cite{bickers}.

This procedure is superior to the integral of the spectral density of the dot level, and is free from
the shortcomings of this density for positive $E_d$, like a spurious peak at the Fermi energy.
This technical point is discussed in Ref. \onlinecite{costi}.

For $N$=2 we include in the figure results obtained using the Bethe ansatz (BA)  as described in 
Ref. \onlinecite{ba}. These results were shifted to the left by a constant energy $C$ to compensate
by the Haldane shift and possible uncertainties in the position of $E_d$ in the BA treatment.
We have used $C=1.67$. We also include in the figure the noninteracting results.

For $N=2$ one can observe a very good agreement between NCA and BA results for all $E_d$.
For positive $E_d$ both results agree also with the $U=0$ (noninteracting) case. However, 
for negative $E_d$, the effects of correlations become apparent. In particular for 
$U=0$ and large negative $E_d$, $n=N$ for the SU($N$) model, while for $U, -E_d \rightarrow \infty$, $n=1$.
This is due to the decreasing weight of the charge-transfer peak, discussed in Section \ref{vari} 
[see also Eq. (\ref{w}) below].

The deviations form the noninteracting case are stronger in the SU(4) case. They are
significant even for $E_d$. This is due in part to the larger Haldane shift in the SU(4) 
case. The fact that the increase in $n$ with decreasing $E_d$ is smoother than for the N=2 case, 
is due to the larger width of the charge-transfer peaks in the region of negative $E_d$.

\section{Possible experiments}

\label{exp}

\subsection{Two dots independently connected to its own pair of leads}

\label{2dots}

In the setup with two dots studied in Refs. \onlinecite{ama,hubel,keller},
the dot $i=1,2$ is connected to left and right ($\nu=L, R$) leads by couplings 
$\Delta_{\nu i}$ and all applied voltages and couplings can be controlled with high accuracy,
For large enough repulsion between the dots,\cite{nishi2} if
$\Delta_{L 1}+\Delta_{R 1}=\Delta_{L 2}+\Delta_{R 2}$, and equal energy levels $E_{d 2}=E_{d 1}$, 
the system is in the SU(4) regime.\cite{keller}
Even if the equality is not satisfied exactly, the SU(4) symmetry can be restored at low energies adjusting 
the difference $E_{d 2} - E_{d 1}$.\cite{restor,nishi}.
In this conditions, if in addition the couplings of one dot are very asymmetric 
(say $\Delta_{L 1} \ll \Delta_{R 1}$), moving the bias voltage $V_{L 1}$ of the $L 1$ lead, the resulting 
differential current through dot 1, $dI_1/dV_{L 1}$ just maps the spectral density of this dot
(as in scanning tunneling spectroscopy).\cite{oks}
In this case, a marked asymmetry in the $dI_1/dV_{L 1}$ response for positive and negative 
$E_{d i}$ should be measured, as in Fig. (\ref{anchos}).

Previous non-equilibrium calculations show that if $\Delta_{L 1} < \Delta_{R 1}/9$, the resulting $dI_1/dV_{L 1}$
is very similar to the equilibrium spectral density for coupling $\Delta_{L 1} + \Delta_{R 1}$ 
and chemical potential corresponding to the right lead of dot 1.

\subsection{Conductance through an SU(4) "impurity" including capacitance
effects}

\label{haugsu4}

Here we discuss an experiment like that of K\"{o}nemann \textit{et al. }\cite{haug} 
introduced in Section \ref{intro}, in which asymmetries with the sign
of bias voltage $V$ were observed in the widths and intensities of
conductance peaks, but here we assume the QD or impurity described by the SU(4) IAM
instead of the SU(2) one discussed previously.\cite{haug,capac} 
Specifically, the interacting system, which we
call impurity is coupled to two SU(4) interacting leads (left $L$ and right $R$) 
with chemical potentials $\mu _{\nu }$ and tunnel couplings $\Delta
_{\nu }$ ($\nu =L$, $R$), as described by the Hamiltonian Eq. (\ref{ham}).

\begin{figure}[h]
\includegraphics[width=8.cm]{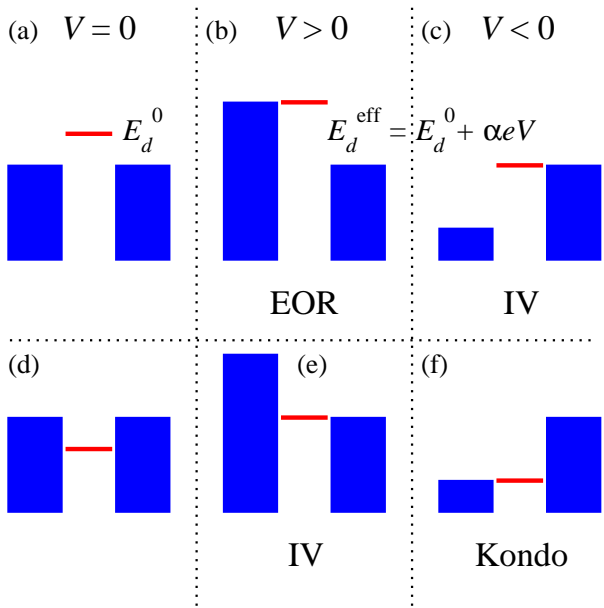}
\caption{(Color online) Scheme of the Fermi level of both leads and the displacement of the dot level
$E-d$ due to capacitance effects. The top (bottom) figures represent different situations for 
positive (negative) $E_{d}^{0}$. When the Fermi level of one of the leads coincides with $E_d$ 
a peak in the conductance is observed. Assuming that the right lead is more strongly coupled 
to the dot, the system can be in Kondo, intermediate valence (IV) or empty orbital regime (EOR)
as indicated.}
\label{scheme}
\end{figure}

We take the sign of the bias voltage in such a way that $\mu _{L}-\mu_{R}=eV $. 
Note that the interchange of right and left leads or electrons by
holes is equivalent to a change of sign of $V$. In this section, we take the
origin of energies at $\mu _{R}=0$. The capacitance effects modify the
energy necessary to add an electron to the dot with the lever arm parameter $\alpha $ 
(which depends on the source, drain and gate capacitances).\cite{haug,park}
Usually $E_{d}=E_{d}^{0}+\alpha eV$ is assumed, where $\alpha \sim 1/2$, so that
$E_{d}$ is displaced in approximately half the magnitude of $\mu _{L}-\mu_{R}$.
Since the experimentally accessible quantity is $E_{d}^{\text{eff}}$
and not $E_{d}$, we assume
\begin{equation}
E_{d}^{\text{eff}}=E_{d}^{0}+\alpha eV,  \label{Ed}
\end{equation}
where $E_{d}^{0}$ is the effective dot level (the position of the CT peak) when $V=0$
and $\alpha$ with $ \le \alpha \le 1$ describes how the effective level is modified by the bias voltage. 
A scheme of how $E_d$ is modified applying a bias voltage is presented in Fig. \ref{scheme}

The current through the impurity is given by \cite{meir}
\begin{eqnarray}
I &=&C\int d\omega \rho _{m}(\omega ,V,E_{d})[f_{L}(\omega )-f_{R}(\omega )],
\notag \\
C &=&\frac{4N\pi e\Delta _{L}\Delta _{R}}{h\Delta },  \label{i}
\end{eqnarray}
where $N=4$ in this section, $f_{\nu }(\omega )=f(\omega -\mu _{\nu })$ is
the Fermi distribution in each lead, with $f(\omega )=1/(e^{\omega /kT}+1)$,
and $\rho _{m}(\omega ,V,E_{d})$ is the non-equilibrium spectral density of
the impurity level with symmetry $m$, which depends on the voltage $V$
explicitly and also implicitly through the voltage dependence of $E_{d}$
[Eq. (\ref{Ed})]. The conductance is $G=dI/dV.$

As shown before,\cite{capac} to observe the asymmetry in the peaks in $G(V)$
one needs asymmetric coupling to the leads. Therefore we assume $\Delta
_{R}\gg \Delta _{L}$. 
It has been shown before \cite{oks} that for $\Delta
_{L} \lesssim \Delta _{R}/9$, the spectral density at the dot practically coincides with the 
corresponding one with the dot at equilibrium with the right lead
at chemical potential $\mu _{R}$. This allows us to avoid a cumbersome
non-equilibrium calculation.

To simplify the calculation further, we extend a phenomenological approach
used before \cite{capac} to the present case, and assume that the impurity
spectral density near the CT peak can be approximated by 
\begin{eqnarray}
\rho _{m}(\omega ) &=&\frac{(w \Delta _{\text{CT}})/\pi 
}{(\omega -E_{d}^{\text{eff}})^{2}+\Delta _{\text{CT}}^{2}},  \label{rf} \\
w &=&1-\sum_{m^{\prime }\neq m}\langle n_{m^{\prime }}\rangle   \label{w}
\end{eqnarray}%
where $\langle n_{m}\rangle =\int d\omega \rho _{m}(\omega )f_{R}(\omega )$
is the occupation of the dot for symmetry $m$.
The weight $w$ is the probability that the site of the impurity is not occupied 
by electrons of symmetry different than $m$. This weight appears naturally in the atomic 
limit $ V_q \rightarrow 0$ and was used in Hubbard-III \cite{hub3}
and similar \cite{aaa,dyn} approximations to Hubbard and intermediate-valence models. 
Integrating Eq. (\ref{rf}), using Eq. (\ref{w}) and SU($N$) symmetry so that $\langle n_{m}\rangle $ is
independent of $m$, one obtains a linear equation for $\langle n_{m}\rangle $. 
Solving it we obtain in general  
\begin{eqnarray}
\langle n_{m}\rangle  &=&\frac{1-2\Psi _{R}}{(N+1)-2(N-1)\Psi _{R}}, 
\notag \\
\Psi _{\nu } &=&\frac{1}{\pi }\text{Im}\psi (\chi _{v}),  \notag \\
\chi _{v} &=&\frac{1}{2}+\frac{\Delta _{\text{CT}}+i(E_{d}^{\text{eff}}-\mu
_{\nu })}{2\pi T},  \label{nm}
\end{eqnarray}%
where $\psi (x)$ is the digamma function.

\begin{figure}[h]
\vspace{1.cm}
\includegraphics[width=8.cm]{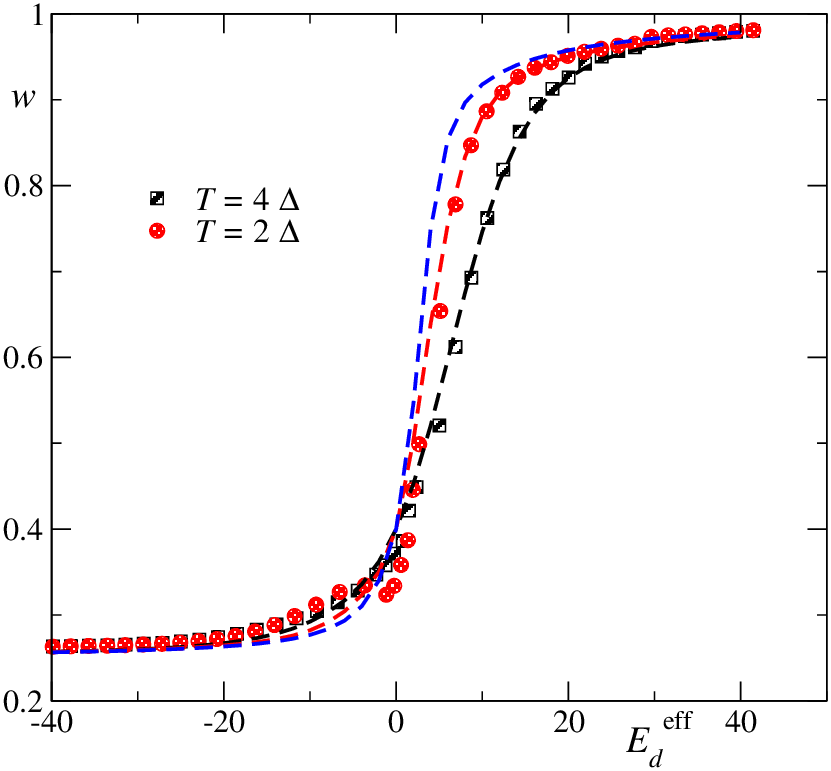}
\caption{(Color online) Weight of the CT peak obtained from the fit of the NCA CT peak (circles and squares) 
and from Eqs. (\ref{w}) and (\ref{nm}) (dashed lines) with $N=4$ as a function of the effective impurity level for
several temperatures $T$.
Blue dashed line corresponds to $T=0$.}
\label{weight}
\end{figure}

In Fig.~\ref{weight} we compare the weight obtained fitting the NCA results with those
of the phenomenological approach given here. With the exception of the region in which both 
$T$ and $E_{d}^{\text{eff}}$ are near zero, we see that Eq. \ref{rf} reproduces well
the NCA result. In particular for $|E_{d}^{\text{eff}}| \gg \Delta$, Eq. (\ref{rf}) 
reproduces accurately the spectral density at the CT peak.

Replacing Eq. (\ref{rf}) in Eq. (\ref{i}) gives the current 
\begin{equation}
I=C [1-(N-1)\langle n_{m}\rangle ](\Psi _{R}-\Psi _{L}),  \label{if}
\end{equation}

Differentiating this expression with respect to the voltage, using Eqs. (\ref{dfit}), (\ref{Ed}),  
and (\ref{nm}), we obtain for $N=4$

\begin{eqnarray}
G(V) &=&C [ \left( 1-3\langle n_{m}\rangle \right) \left( \Psi
_{R}^{\prime }-\Psi _{L}^{\prime }\right) \notag \\
&&+\frac{12\Psi _{R}^{\prime }}{\left( 5-6\Psi _{R}\right) ^{2}}(\Psi _{R}-\Psi _{L})] , \notag \\ 
\Psi _{L}^{\prime } &=&\frac{1}{2\pi ^{2}T}\text{Im}\left\{ \psi ^{\prime
}(\chi _{L})\left( i(\alpha -1)-\xi \right) \right\} , \notag \\
\Psi _{R}^{\prime } &=&\frac{1}{2\pi ^{2}T}\text{Im}\left\{ \psi ^{\prime
}(\chi _{R})\left( i\alpha -\xi \right) \right\} , \notag \\
\xi  &=&\frac{b\alpha }{c}\text{Sech}\left[ \frac{E_{d}^{\text{eff}}
-E_{\text{IV}}}{c\Delta }\right] ^{2},
\end{eqnarray}
and $\psi ^{\prime }(x)$ is the derivative of the digamma function.

The weak point in this approach is the use of the fit Eq. (\ref{dfit}), which as can
be seen in Figs. \ref{anchos} and \ref{anchost} fails near the intermediate valence regime ($E_{d}^{\text{eff}} \sim 0$).
However, this equation is accurate enough in the Kondo ($-E_{d}^{\text{eff}} \gg N \Delta$) and
empty orbital ($E_{d}^{\text{eff}} \gg \Delta$) regimes.

\begin{figure}[h]
\vspace{1.cm}
\includegraphics[width=8.cm]{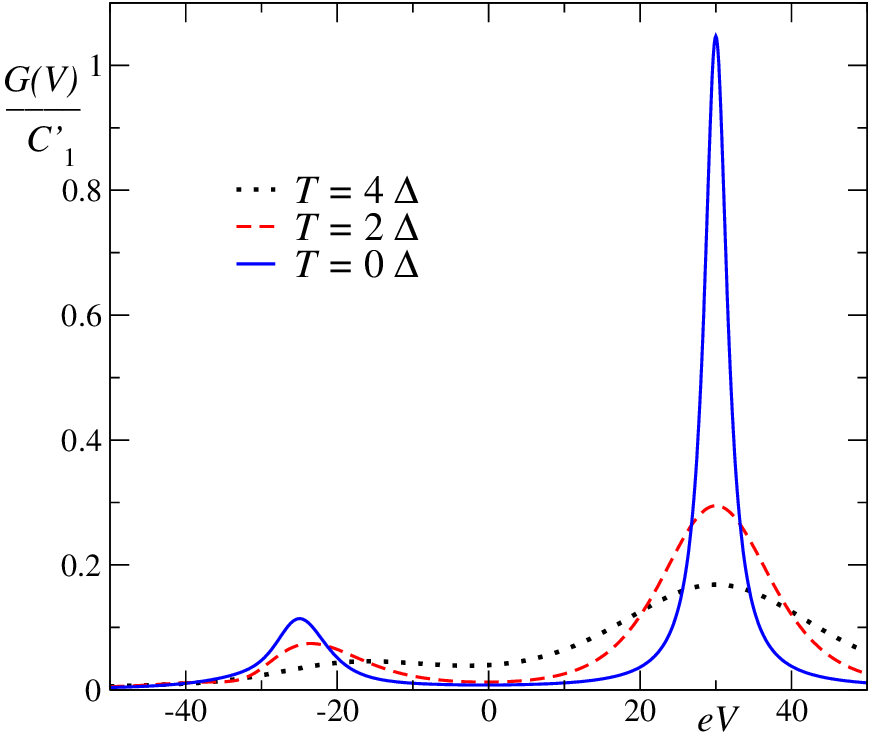}
\caption{(Color online) Differential conductance for an SU(4) impurity as a function of bias voltage
for asymmetric couplings $\Delta_R \gg \Delta_L$, capacitance ratio $\alpha=1/2$, 
$E_{d}^0=15 \Delta$
and several temperatures $T$. The constant $C^{\prime}_{1} = eC/(2 \pi \Delta)$}
\label{haugp}
\end{figure}

In Fig.~\ref{haugp} we show the resulting conductance for $E_{d}^0$ slightly smaller than
$E_{d}^{\text{eff}}$ of the dotted line in Fig.~\ref{nca}. 
At $V=0$, $E_{d}^{\text{eff}}$ lies $15 \Delta$ at higher energy 
than $\mu_L=\mu_R=0$. as $V$ increases, a peak develops which reaches its maximum at $eV= 30 \Delta$, for which
$\mu_L=E_{d}^{\text{eff}}=eV$. This configuration is similar to a scanning tunneling microscope (STM), for which the tip
is very weakly coupled to the impurity. However, due to capacitance effects, the peak is broader by a factor 
$1/(1-\alpha)$ (two in the figure),\cite{capac} and the intensity at the maximum is near $(1-\alpha)eC/(\pi \Delta)$  
as it can be seen in the figure. This peak, which corresponds to the empty orbital regime, 
is accurately reproduced by our phenomenological approach.
When a negative voltage is applied, $E_{d}^{\text{eff}}$ decreases until for $eV=-30 \Delta$ it reaches the value 
of $\mu_R=0$. In this case, the system is in the intermediate-valence regime and for $T=0$ our approach is not accurate because 
of the uncertainty in the width and weight of the CT in this region. From the spectral density, one would expect 
a peak nearly two times wider and a total weight reduced by a factor 2/5, but also the strong dependence of
$\Delta_{\text{CT}}$ on $V$ (through $E_{d}^{\text{eff}}$) affects the shape. 

As expected, increasing the temperature has the effect of broadening both peaks, with more dramatic effects on 
the narrower one. 

\begin{figure}[h]
\vspace{1.cm}
\includegraphics[width=8.cm]{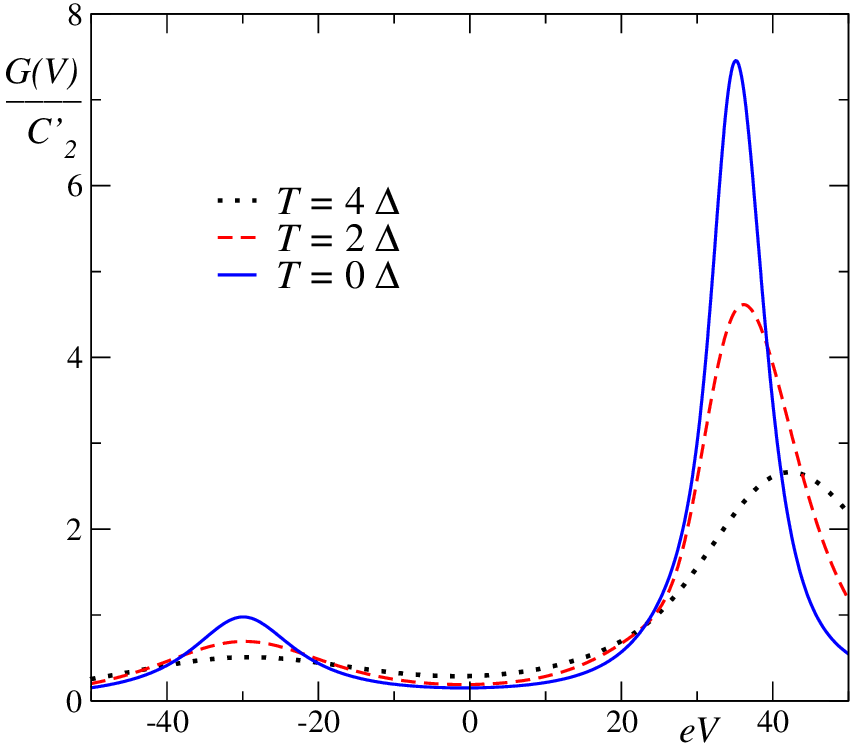}
\caption{(Color online) Same as Fig. \ref{haugp} for $E_{d}^0=-15 \Delta$. The constant 
$C^{\prime}_{2} = eC/(32 \pi \Delta)$}
\label{haugn}
\end{figure}

If the system is instead prepared starting from the Kondo regime $E_{d}^0=-15 \Delta$ (Fig.~\ref{haugn}),
at low temperatures one has the well known Kondo peak near zero bias.\cite{capac} Applying a positive $V$ 
one reaches again the intermediate-valence regime for $eV= 30 \Delta$ for which $E_{d}^{\text{eff}}=\mu_R=0$. 
Applying instead a negative voltage $eV \sim -30 \Delta$,  $\mu_L=eV$ again coincides with $E_{d}^{\text{eff}}$ and 
a peak corresponding to a configuration similar to an STM one, but with a width increased by a factor 
$1/(1-\alpha)$ with respect to the spectral density, is obtained. Due to the increase of the width by a factor 
$N$ and a decrease in the total weight of the peak by a factor $\sim 1/N$, the intensity at this relative maximum 
in $G(V)$ is near $(1-\alpha)eC/(N^2\Delta)$, as seen in the figure.

This peak corresponds to the Kondo regime and is well 
reproduced by our approach. Then we can anticipate that if an experiment with symmetric values of $E_{d}^0$ (with the same
$|E_{d}^0|$ can be achieved, applying a gate voltage to the less coupled lead with the sign such that
$|E_{d}^0|$ increases, in the case of negative $E_{d}^0$ (Kondo regime) the peak will be four times wider and nearly
sixteen times less intense than for positive $E_{d}^0$ (empty orbital regime).

\section{Summary and discussion}

\label{sum}

We have calculated the width of the charge transfer peak in the infinite-$U$
SU(4) Anderson model as a function of the impurity level $E_d$ and 
for the general SU($N$) case in the Kondo regime
$-E_d \gg N \Delta$, where $\Delta$ is half the resonant level width.
While it is clear that in the empty orbital regime $E_d \gg  \Delta$,
the half width at half maximum of the charge transfer peak
$\Delta_{\text{CT}}= \Delta$, we obtain that in the Kondo regime
$\Delta_{\text{CT}}=N\Delta$ extending previous results for the SU(2) case.\cite{pru,logan}
The weight of the peak in the Kondo regime is reduced by a factor $\sim 1/N$ due
correlation effects, so that the maximum intensity is reduced by a factor $1/N^2$.
The variational calculation presented in Section \ref{vari} provide
a simple physical picture of the non-trivial effect of correlations in
broadening the charge transfer peak in the Kondo regime. 

The variational calculation of Section \ref{vari} suggest a simple interpretation
of the width $\Delta_{\text{CT}}$. For a peak below the Fermi level $\epsilon_F=0$,
the shape of the spectral density is defined by the spectral distribution of the 
state $|v>$ (not an eigenstate) obtained after 
annihilating one $d_m$ electron [see the Eq. (\ref{rhov}) for $\rho _{m}^{d}(\omega )$]. 
The hybridization term mixes this state with other states in which 
a $d_m$ electron is created again. The number of ways $M$ of adding this electron 
gives the number of bands that hybridize with $|v>$ and the width would be
$\Delta_{\text{CT}}=M \Delta$. 
For peaks above the Fermi energy the same argument would apply interchanging creation and annihilation.
To check this idea and investigate the effects of
finite $U$, we have run the DDMRG code for $E_d=-30$, and $U=20$. We obtain peaks 
at $E_d+U$ and $E_d+2U$ of width $\sim 3.2 \Delta$ consistent with the expected 
$M=3$. However, there is also an intense central peak which overlaps 
with the charge-transfer peak and renders it dangerous to draw definite conclusions.

To analyze general values of $E_d$, it is more useful to discuss it
in terms of the position of the charge-transfer peak $E_{d}^{\text{eff}}$
which incorporates a shift in the effective $E_d$ discussed first by Haldane for
the SU(2) case and more recently for the SU(4) case.\cite{fili,restor}.
Our results for SU(4) symmetry suggest a rather abrupt decrease of $\Delta_{\text{CT}}$ as 
$E_{d}^{\text{eff}}$ increases from $\sim -\Delta$ to  $\sim \Delta$, but 
our analysis at low temperature for small $|E_{d}^{\text{eff}}|$ is complicated by the merge 
of the charge-transfer and Kondo peaks and limitations of our approaches. 

The strong contrast between the charge-transfer peaks in the Kondo and empty-orbital
regimes should be observable in experiment. This is discussed in Section \ref{exp}.
One possibility is a setup of two quantum dots,\cite{ama,hubel,keller} which
can be tuned to prepare the system in a similar way as in scanning tunneling spectroscopies,
with one quantum dot very weakly coupled to a lead to which a bias voltage is applied.
Another possibility is to extend to an SU(4) system the experiment of
K\"{o}nemann \textit{et al.}, in which strong asymmetries were observed 
in the width and magnitude of conductance peaks as a function of gate voltage
for a quantum dot described by the SU(2) impurity Anderson model.\cite{haug}
These experiments, in which also capacitance effects play a role,\cite{capac}
would display stronger asymmetries in the SU(4) case.

We hope that these results will encourage experimental work along these lines 
and be useful for the interpretation of Coulomb blockade peaks in conductance experiments.
On the theoretical side, a deeper analysis of the change in $\Delta_{\text{CT}}$
at zero temperature 
as $E_{d}^{\text{eff}}$ crosses zero would be useful.

\section*{Acknowledgments}

J. F. and A. A. A. are sponsored by PIP 112-201101-00832 of CONICET and PICT
2013-1045 of the ANPCyT. F. L.  and C. G.  acknowledge support from CONICET, through Project PIP
No. 112-201201-00389CO


\begin{thebibliography}{99}
\bibitem{hewson} A. C. Hewson, \textit{The Kondo Problem to Heavy Fermions}
(Cambridge University Press, Cambridge, England, 1997), ISBN 9780521599474.

\bibitem{gold} \textit{Kondo effect in a single-electron transistor}, D.
Goldhaber-Gordon, H. Shtrikman, D. Mahalu, D. Abusch-Magder, U. Meirav, and
M. A. Kastner, Nature \textbf{391}, 156 (1998).

\bibitem{cro} \textit{A Tunable Kondo Effect in Quantum Dots}, S. M.
Cronenwett, T. H. Oosterkamp, and L. P. Kouwenhoven, Science \textbf{281},
540 (1998).

\bibitem{gold2} \textit{From the Kondo Regime to the Mixed-Valence Regime in
a Single-Electron Transistor}, D. Goldhaber-Gordon, J. G\"{o}res, M. A.
Kastner, H. Shtrikman, D. Mahalu, and U. Meirav, Phys. Rev. Lett. \textbf{81}, 5225 (1998).

\bibitem{wiel} \textit{The Kondo Effect in the Unitary Limit}, W.G. van der
Wiel, S. de Franceschi, T. Fujisawa, J.M. Elzerman, S. Tarucha, and L.P.
Kowenhoven, Science \textbf{289}, 2105 (2000).

\bibitem{grobis} \textit{Universal Scaling in Nonequilibrium Transport
through a Single Channel Kondo Dot}, M. Grobis, I. G. Rau, R. M. Potok, H.
Shtrikman, and D. Goldhaber-Gordon, Phys. Rev. Lett. \textbf{100}, 246601
(2008).

\bibitem{kreti} \textit{Spin-$\frac{1}{2}$ Kondo effect in an InAs nanowire
quantum dot: Unitary limit, conductance scaling, and Zeeman splitting}, A.
V. Kretinin, H. Shtrikman, D. Goldhaber-Gordon, M. Hanl, A. Weichselbaum, J.
von Delft, T. Costi, and D. Mahalu, Phys. Rev. B \textbf{84}, 245316 (2011).

\bibitem{ama} \textit{Pseudospin-Resolved Transport Spectroscopy of the
Kondo Effect in a Double Quantum Dot}, S. Amasha, A. J. Keller, I. G. Rau,
A. Carmi, J. A. Katine, H. Shtrikman, Y. Oreg, and D. Goldhaber-Gordon,
Phys. Rev. Lett. \textbf{110}, 046604 (2013).

\bibitem{hubel} \textit{Correlated Electron Tunneling through Two Separate 
Quantum Dot Systems with Strong Capacitive Interdot Coupling}, 
A. H\"ubel, 
K. Held, J. Weis, and K. v. Klitzing, Phys. Rev. Lett. \textbf{101}, 186804 (2008).

\bibitem{keller} \textit{Emergent SU(4) Kondo physics in a spin–charge-entangled 
double quantum dot},
A. J. Keller, S. Amasha, I. Weymann, C. P. Moca, I. G. Rau,
J. A. Katine, H. Shtrikman, G. Zar\'and, and D. Goldhaber-Gordon, Nat. Phys. 
\textbf{10}, 145 (2014).

\bibitem{liang} \textit{Kondo resonance in a single-molecule transistor}, 
W. Liang, M. P. Shores, M. Bockrath, J. R. Long, and H. Park, Nature \textbf{417}, 725 (2002).

\bibitem{kuba} \textit{Single-electron transistor of a single organic
molecule with access to several redox states}, 
S. Kubatkin, A. Danilov, M.
Hjort, J. Cornil, J. L, Br\'edas, N. Stuhr-Hansen, P. Hedeg\aa rd, and Th. Bj%
\o rnholm, Nature \textbf{425}, 699 (2003).

\bibitem{yu} \textit{Kondo Resonances and Anomalous Gate Dependence in the
Electrical Conductivity of Single-Molecule Transistors}, 
L. H. Yu, Z. K.
Keane, J. W. Ciszek, L. Cheng, J. M. Tour, T. Baruah, M. R. Pederson, and D.
Natelson, Phys. Rev. Lett. \textbf{95}, 256803 (2005).

\bibitem{leuen} \textit{Berry-Phase Oscillations of the Kondo Effect in
Single-Molecule Magnets}, 
M. N. Leuenberger and E. R. Mucciolo, Phys. Rev.
Lett. \textbf{97}, 126601 (2006).

\bibitem{parks} \textit{Tuning the Kondo Effect with a Mechanically
Controllable Break Junction}, 
J. J. Parks, A. R. Champagne, G. R. Hutchison,
S. Flores-Torres, H. D. Abru\~{n}a, and D. C. Ralph, Phys. Rev. Lett. 
\textbf{99}, 026601 (2007).

\bibitem{roch} \textit{Quantum phase transition in a single-molecule quantum
dot}, 
N. Roch, S. Florens, V. Bouchiat, W. Wernsdorfer, and F. Balestro,
Nature \textbf{453}, 633 (2008).

\bibitem{scott} \textit{Universal scaling of nonequilibrium transport in the
Kondo regime of single molecule devices}, 
G. D. Scott, Z. K. Keane, J. W.
Ciszek, J. M. Tour, and D. Natelson, Phys. Rev. B \textbf{79}, 165413 (2009).

\bibitem{parks2} \textit{Mechanical Control of Spin States in Spin-1
Molecules and the Underscreened Kondo Effect}, 
J. J. Parks, A. R. Champagne,
T. A. Costi, W. W. Shum, A. N. Pasupathy, E. Neuscamman, S. Flores-Torres,
P. S. Cornaglia, A. A. Aligia, C. A. Balseiro, G. K.-L. Chan, H. D. Abru\~{n}
a, and D. C. Ralph, Science \textbf{328}, 1370 (2010).

\bibitem{NatelsonReview} \textit{Kondo Resonances in Molecular Devices}, 
G. D. Scott and D. Natelson, ACS Nano \textbf{4}, 3560 (2010).

\bibitem{serge} \textit{Universal transport signatures in two-electron
molecular quantum dots: gate-tunable Hund's rule, underscreened Kondo effect
and quantum phase transitions}, 
S. Florens, A, Freyn, N. Roch, W.
Wernsdorfer, F. Balestro, P. Roura-Bas and A. A. Aligia, J. Phys. Condens.
Matter \textbf{23}, 243202 (2011); references therein.

\bibitem{vincent} \textit{Electronic read-out of a single nuclear spin using
a molecular spin transistor}, 
R. Vincent, S. Klyatskaya, M. Ruben, W.
Wernsdorfer, and F. Balestro, Nature (London) \textbf{488}, 357 (2012).

\bibitem{jari} \textit{Electronic Transport Spectroscopy of Carbon Nanotubes in 
a Magnetic Field}, 
P. Jarillo-Herrero, J. Kong, H. S. J. van der Zant, C.
Dekker, L. P. Kouwenhoven, and S. De Franceschi, Phys. Rev. Lett. \textbf{94}
, 156802 (2005).

\bibitem{maka} \textit{SU(2) and SU(4) Kondo effects in carbon nanotube quantum 
dots}, 
A. Makarovski, A. Zhukov, J. Liu, and G. Finkelstein, Phys.
Rev. B \textbf{75}, 241407 (2007).

\bibitem{anders} \textit{Zero-Bias Conductance in Carbon Nanotube Quantum Dots}, 
F. B. Anders, D. E. Logan, M. R. Galpin, and G. Finkelstein, Phys. Rev. Lett. 
\textbf{100}, 086809 (2008).

\bibitem{haug} \textit{Tunneling resonances in quantum dots: Coulomb interaction 
modifies the width}, 
J. K\"onemann, B. Kubala, J. K\"onig, and R. J. Haug, Phys.
Rev. B \textbf{73}, 033313 (2006).

\bibitem{park} J. Park, Ph. D Thesis, University of California (2003).

\bibitem{capac} \textit{Impact of capacitance and tunneling asymmetries on
Coulomb blockade edges and Kondo peaks in nonequilibrium transport through
molecular quantum dots}, 
A. A. Aligia, P. Roura-Bas, and S. Florens, Phys.
Rev. B \textbf{92}, 035404 (2015).

\bibitem{pru} \textit{The Anderson model with finite Coulomb repulsion}, 
Th. Pruschke and N. Grewe, Z. Phys. B \textbf{74}, 439 (1989).

\bibitem{logan} \textit{A local moment approach to the Anderson model}, 
D. E. Logan, M. P. Eastwood, and M. A. Tusch, J. Phys.Condens. Matter 
\textbf{10}, 2673 (1998).

\bibitem{tetta} \textit{Magnetic-Field Probing of an SU(4) Kondo Resonance in 
a Single-Atom Transistor}, 
G. C. Tettamanzi, J. Verduijn, G. P. Lansbergen, M.
Blaauboer, M. J. Calder\'{o}n, R. Aguado, and S. Rogge, Phys. Rev. Lett. 
\textbf{108}, 046803 (2012).

\bibitem{lans} \textit{Tunable Kondo Effect in a Single Donor Atom}, 
G. P. Lansbergen, G. C. Tettamanzi, J. Verduijn, N. Collaert,
S. Biesemans, M. Blaauboer, and S. Rogge, Nano Lett. \textbf{10}, 455 (2010).

\bibitem{restor} \textit{Restoring the SU(4) Kondo regime in a double quantum 
dot system},
L. Tosi, P. Roura-Bas, and A. A. Aligia, J. Phys.: Condens.
Matter \textbf{27}, 335601 (2015).

\bibitem{nishi} \textit{Conditions for observing emergent SU(4) symmetry in a 
double quantum dot}, 
Y. Nishikawa, O. J. Curtin, A. C. Hewson, D. J. G. Crow, and
J. Bauer, Phys. Rev. B \textbf{93}, 235115 (2016).

\bibitem{buss} \textit{Transport in carbon nanotubes: Two-level SU(2) regime reveals 
subtle competition between Kondo and intermediate valence states},
C. A. B\"{u}sser, E. Vernek, P. Orellana, G. A. Lara, E. H.
Kim, A. E. Feiguin, E. V. Anda, and G. B. Martins, Phys. Rev. B \textbf{83},
125404 (2011).

\bibitem{desint} \textit{Interplay between quantum interference and Kondo
effects in nonequilibrium transport through nanoscopic systems}, 
P. Roura-Bas, L. Tosi, A. A. Aligia, and K. Hallberg, Phys. Rev. B \textbf{84},
073406 (2011).

\bibitem{buss2} \textit{Electrostatic control over polarized currents through 
the spin-orbital Kondo effect},
C. A. B\"{u}sser, A. E. Feiguin, and G. B. Martins, Phys. Rev. B \textbf{85}, 
241310(R) (2012).

\bibitem{thermo} \textit{Thermopower of an SU(4) Kondo resonance under an
SU(2) symmetry-breaking field}, 
P. Roura-Bas, L. Tosi, A. A. Aligia, and P.
S. Cornaglia, Phys. Rev. B \textbf{86}, 165106 (2012).

\bibitem{nishi2} \textit{Analysis of low-energy response and possible emergent 
SU(4) Kondo state in a double quantum dot}, 
Y. Nishikawa, A. C. Hewson, D. J.G. 
Crow, and J. Bauer, Phys. Rev. B \textbf{88}, 245130 (2013).

\bibitem{oks} \textit{Orbital Kondo spectroscopy in a double quantum dot system},
 L. Tosi, P. Roura-Bas, and A. A. Aligia, Phys. Rev. B \textbf{88}, 235427 (2013).

\bibitem{lopes} \textit{SU(4)-SU(2) crossover and spin-filter properties of a double 
quantum dot nanosystem}, 
V. Lopes, R. A. Padilla, G. B. Martins, and E. V. Anda
Phys. Rev. B \textbf{95}, 245133 (2017).

\bibitem{klee} \textit{Abrupt disappearance and re-emergence of the SU(4)
and SU(2) Kondo effects due to population inversion}, 
Y. Kleeorin and Y.
Meir, Phys. Rev. B \textbf{96}, 045118 (2017).

\bibitem{varma} \textit{Magnetic susceptibility of mixed-valence rare-earth 
compounds}, 
C. M. Varma and Y. Yafet, Phys. Rev. B \textbf{13}, 2950 (1976).

\bibitem{gunn} \textit{Photoemission from Ce Compounds: Exact Model
Calculation in the Limit of Large Degeneracy}, 
O. Gunnarsson and K.
Sch\"onhammer, Phys. Rev. Lett. \textbf{50}, 604 (1983).

\bibitem{hald} \textit{Scaling Theory of the Asymmetric Anderson Model},
F. D. M. Haldane, Phys. Rev. Lett. \textbf{40}, 416 (1978).


\bibitem{bickers} \textit{Review of techniques in the large-N expansion for dilute 
magnetic alloys}, 
N.E. Bickers, Rev. of Mod. Phys. \textbf{59}, 845 (1987).

\bibitem{kroha} \textit{Fermi and Non-Fermi Liquid Behavior in Quantum Impurity 
Systems: Conserving Slave Boson Theory}, 
J. Kroha and P. W\"{o}lfle, Acta Phys. Pol. B \textbf{29}, 3781 (1998).


\bibitem{zitko} \textit{Quantitative determination of the discretization and 
truncation errors in numerical renormalization-group calculations of spectral 
functions}, 
R. \v{Z}itko, Phys. Rev. B \textbf{84}, 085142 (2011).

\bibitem{loig} \textit{Spectral density of an interacting dot coupled indirectly 
to conducting leads}, 
L. Vaugier, A.A. Aligia and A.M. Lobos, Phys. Rev. B \textbf{%
76}, 165112 (2007).


\bibitem{ds} \textit{Kondo temperature when the Fermi level is near a step in the 
conduction density of states}, 
J. Fern\'{a}ndez, A. A. Aligia, P. Roura-Bas, and J. A. Andrade,
Phys. Rev. B 95, 045143 (2017).

\bibitem{wingreen} \textit{Anderson model out of equilibrium: Noncrossing-approximation 
approach to transport through a quantum dot}, 
N. S. Wingreen and Y. Meir, Phys. Rev. 
B \textbf{49}, 11040 (1994).

\bibitem{reinert} \textit{Temperature Dependence of the Kondo Resonance and Its Satellites in CeCu$_{2}$Si$_{2}$}, 
F. Reinert, D. Ehm, S. Schmidt, 
G. Nicolay, S. H\"{u}fner, J. Kroha, O. Trovarelli, and C. Geibel, Phys. Rev. Lett. 
\textbf{87}, 106401 (2001).

\bibitem{ehm} \textit{High-resolution photoemission study on low-TK Ce systems: 
Kondo resonance, crystal field structures, and their temperature dependence}, 
D. Ehm, S. H\"{u}fner, F. Reinert, J. Kroha, P. W\"{o}lfle, O.Stockert, C. Geibel,
 and H. v. L\"{o}hneysen, Phys. Rev. B \textbf{76}, 045117 (2007).
 
\bibitem{fon} \textit{Nonequilibrium transport through magnetic
vibrating molecules}, Phys. Rev. B \textbf{87}, 195136 (2013) P. Roura-Bas,
L. Tosi and A. A. Aligia.

\bibitem{sate} \textit{Replicas of the Kondo peak due to electron-vibration interaction in molecular 
transport properties},
P. Roura-Bas, L. Tosi, and A. A. Aligia
Phys. Rev. B \textbf{93}, 115139 (2016).

\bibitem{NFL} \textit{Non-Fermi-liquid behavior in nonequilibrium transport through 
Co-doped Au chains connected to fourfold symmetric leads}, 
S. Di Napoli, P. Roura-Bas, 
A. Weichselbaum and A. A. Aligia, Phys. Rev. B \textbf{90}, 125149 (2014).

\bibitem{st} \textit{Nonequilibrium transport through a singlet-triplet Anderson 
impurity}, 
P. Roura Bas and A. A. Aligia, Phys. Rev. B \textbf{80}, 035308
(2009); J. Phys. Condens. Matt. \textbf{22}, 025602 (2010).

\bibitem{hbo} \textit{Non-equilibrium differential conductance through a quantum 
dot in a magnetic field}, 
A. C. Hewson, J. Bauer, and A. Oguri, J. Phys.: Condens.
Matter \textbf{17}, 5413 (2005), and references therein.


\bibitem{ng} \textit{Nonequilibrium self-energies, Ng approach, and heat current 
of a nanodevice for small bias voltage and temperature}, 
A. A. Aligia, Phys. Rev. 
B \textbf{89}, 125405 (2014), and references therein.

\bibitem{com3} \textit{Comment on "The renormalized superperturbation theory
(rSPT) approach to the Anderson model in and out of equilibrium"}, 
A. A. Aligia, arXiv:1706.06029.

\bibitem{rome1} \textit{Spin fluctuation effects on the conductance through
a single Pd atom contact}, 
M. A. Romero, S C G\'omez-Carrillo, P. G,
Bolcatto, and E. C. Goldberg, J. Phys. Condens. Matter \textbf{21}, 215602
(2009).

\bibitem{rapha} \textit{Anderson model out of equilibrium: Decoherence
effects in transport through a quantum dot}, 
R. Van Roermund, S. Y. Shiau, and M. Lavagna Phys. Rev. B \textbf{81}, 165115 (2010).

\bibitem{rome2} \textit{Effective treatment of charge and spin fluctuations
in dynamical and static atom-surface interactions}, 
M. A. Romero, F. Flores,
and E. C. Goldberg, Phys. Rev. B \textbf{80}, 235427 (2009).


\bibitem{kuhne99a} \textit{Dynamical correlation functions using the density matrix renormalization group}, 
T.~D. K\"uhner and S.~R. White, Phys. Rev. B {\bf 60},  335  (1999).

\bibitem{hovel00} T. H\"ovelborn, diploma thesis, Bonn/K\"oln, 2000;
available at www.thp.uni-koeln.de/$\widetilde{\phantom{w}}${}gu.

\bibitem{freun93a} \textit{A Transpose-Free Quasi-Minimal Residual Algorithm 
for Non-Hermitian Linear Systems},
R.~W. Freund, SIAM J. Sci. Comput. {\bf 14}, 470 (1993).

\bibitem{jecke02} \textit{Dynamical density-matrix renormalization-group method}, 
E. Jeckelmann, Phys. Rev. B {\bf 66},  045114  (2002).

\bibitem{alva} \textit{Spectral Functions with the Density Matrix Renormalization Group:
Krylov-space Approach for Correction Vectors}, 
A. Nocera and G. Alvarez, Phys. Rev. E {\bf 94}, 053308 (2016).

\bibitem{Gebhard2003} \textit{Fourth-order perturbation theory for the half-filled 
Hubbard model in infinite dimensions}, 
F. Gebhard, E. Jeckelmann, S. Mahlert, S. Nishimoto,
and R. M. Noack, Eur. Phys. J. B {\bf 36}, 491 (2003).
%, ISSN 1434-6036, URL http://dx.doi.org/10.1140/epjb/e2004-00005-5.

\bibitem{Ulbricht2010} \textit{Tracking spin and charge with spectroscopy 
in spin-polarised 1D systems}, 
T. Ulbricht and P. Schmitteckert, EPL {\bf 89}, 47001 (2010).
%, URL http://stacks.iop.org/0295-5075/89/i=4/a=47001.

\bibitem{Weichselbaum2009} \textit{Variational matrix-product-state approach to 
quantum impurity models}, 
A. Weichselbaum, F. Verstraete, U. Schollw\"ock, J. I.
Cirac, and J. von Delft, Phys. Rev. B 80, 165117 (2009).
%,URL http://link.aps.org/doi/10.1103/PhysRevB.80.165117.

\bibitem{Nishimoto2004} \textit{Density-matrix renormalization group approach to 
quantum impurity problems}, 
S. Nishimoto and E. Jeckelmann, J. Phys.: 
Condens. Matter {\bf 16}, 613 (2004).
%, URL http://stacks.iop.org/0953-8984/16/i=4/a=010.

\bibitem{Raas2005} \textit{Spectral densities from dynamic density-matrix 
renormalization}, 
C. Raas and S. G. Uhrig, Eur. Phys. J. B {\bf 45}, 293 (2005).
%, ISSN 1434-6036, URL http://dx.doi.org/10.1140/epjb/e2005-00194-3.

\bibitem{Raas2004} \textit{High Energy Dynamics of the Single Impurity Anderson Model}, 
C. Raas, G. S. Uhrig, and F. B. Anders, Phys. Rev. B {\bf 69}, 041102 (2004).
%, URL http://link.aps.org/doi/10.1103/PhysRevB.69.041102.

\bibitem{Paech1014} \textit{Blind deconvolution of density-matrix 
renormalization-group spectra}, M. Paech and E. Jeckelmann, Phys. Rev. B {\bf 89},
195101 (2014).
%, URL http://link.aps.org/doi/10.1103/PhysRevB.89.195101.

\bibitem{fili} \textit{Kondo temperature of SU(4) symmetric quantum dots},
M. Filippone, C. P. Moca, G. Zar\'{a}nd, and C. Mora, Phys. Rev. B \textbf{90}, 121406(R) (2014).

\bibitem{note} In the DDMRG calculations the two chains were attached to two 
semi-infinite chains of conducting electrons destribed by a tight binding Hamiltonian 
with hopping $t=D/2$. In this case $\Delta(\omega)$ has an energy dependence given by 
$\Delta(\omega)=\Delta [1-(\omega/D)^2$.\cite{proe} For $D=100$ as we have taken, 
this dependence is weak. For example $\Delta(\omega)=0.92 \Delta$ for $|\omega=40|$.

\bibitem{proe} \textit{Kondo and anti-Kondo resonances in transport through nanoscale devices},
A. A. Aligia and C.R. Proetto, Phys. Rev. B \textbf{65}, 165305 (2002)

\bibitem{lang} \textit{Friedel Sum Rule for Anderson's Model of Localized Impurity States},
Phys. Rev. \textbf{150}, 516 (1966).

\bibitem{yoshi} \textit{Restricted Friedel sum rules and Korringa relations
as consequences of conservation laws},
A Yoshimori and A Zawadowski, J. Phys. C \textbf{15}, 5241 (1982).

\bibitem{costi} \textit{Spectral properties of the Anderson impurity model: 
Comparison of numerical-renormalization-group and noncrossing-approximation results}
Costi, T.A., J. Kroha, and P. W\"olfle, Phys. Rev. B \textbf{53}, 1850 (1996).

\bibitem{ba} \textit{Self-consistent hybridization expansions for static properties of the Anderson impurity model},
I. J. Hamad, P. Roura-Bas, A. A. Aligia, and E. V. Anda,  
Physica Status Solidi (b) \textbf{253}, 478 (2015).

\bibitem{meir} \textit{Landauer formula for the current through an interacting 
electron region}, 
Y. Meir and N. S. Wingreen, Phys. Rev. Lett. \textbf{68},
2512 (1992).

\bibitem{hub3} \textit{Electron correlations in narrow energy bands III. 
An improved solution}, 
J. Hubbard, Proc. Roy. Soc. \textbf{A281} (1964) 401.

\bibitem{aaa} \textit{Self-consistent alloy treatment of the periodic anderson 
model: Susceptibility and specific heat of intermediate valence compounds}, 
H. J. Leder and G. Czycholl, Z. Phys. B. \textbf{35}, 7 (1979).

\bibitem{dyn} \textit{Dynamical magnetic susceptibility of intermediate 
valence Tm systems}, 
A. A. Aligia and B. Alascio, J. Magn. Magn. Mat. \textbf{46},
321 (1985).


\end{thebibliography}
\end{document}